\newcommand{\NPA}[3]{Nucl.\ Phys.\ {\bf A#1},\ #2 (#3)}
\newcommand{\NPB}[3]{Nucl.\ Phys.\ {\bf B#1},\ #2 (#3)}

\newcommand{\PLB}[3]{Phys.\ Lett.\ B\ {\bf #1},\ #2 (#3)}

\newcommand{\PRL}[3]{Phys.\ Rev.\ Lett.\ {\bf #1},\ #2 (#3)}

\newcommand{\PRC}[3]{Phys.\ Rev.\ C\ {\bf #1},\ #2 (#3)}
\newcommand{\PRD}[3]{Phys.\ Rev.\ D\ {\bf #1},\ #2 (#3)}

\newcommand{\diracslash}[1]{#1\llap{/\kern2pt}}

\newcommand{\be}{\begin{equation}}
\newcommand{\ee}{\end{equation}}
\newcommand{\bea}{\begin{eqnarray}}
\newcommand{\eea}{\end{eqnarray}}
\newcommand{\ba}[1]{\begin{array}{#1}}
\newcommand{\ea}{\end{array}}

\documentclass{elsart1p}
\usepackage{epsfig,graphicx,pstricks}
\usepackage{psfrag}
\usepackage{color}
\usepackage{amsmath}
\usepackage{amsfonts}
\usepackage{amssymb}
\usepackage{textcomp}
\usepackage{multirow}

\begin{document}
\begin{frontmatter}
% notes:
% use the corauthref command within \author for corresponding author
% footnotes;
\title{Chiral symmety breaking in 3-flavor Nambu-Jona Lasinio model in magnetic background}
\author[Ahmedabad]{Bhaswar Chatterjee}
\author[Ahmedabad]{Hiranmaya Mishra}
\author[Delhi]{Amruta Mishra}
\address[Ahmedabad]{Theory Division, Physical Research Laboratory,
Navrangpura, Ahmedabad 380 009, India}
\address[Delhi]{Department of Physics, Indian Institute of Technology, New 
Delhi-110016,India}

\def\be{\begin{equation}}
\def\ee{\end{equation}}
\def\bearr{\begin{eqnarray}}
\def\eearr{\end{eqnarray}}
\def\zbf#1{{\bf {#1}}}
\def\bfm#1{\mbox{\boldmath $#1$}}
\def\hf{\frac{1}{2}}
\def\sl{\hspace{-0.15cm}/}
\def\omit#1{_{\!\rlap{$\scriptscriptstyle \backslash$}
{\scriptscriptstyle #1}}}
\def\vec#1{\mathchoice
        {\mbox{\boldmath $#1$}}
        {\mbox{\boldmath $#1$}}
        {\mbox{\boldmath $\scriptstyle #1$}}
        {\mbox{\boldmath $\scriptscriptstyle #1$}}
}

\begin{abstract}
Effect of magnetic field on chiral symmetry breaking in a 3-flavor Nambu Jona Lasinio (NJL)
model at finite temperature and densities is considered here using an explicit structure 
of ground state in terms of quark and antiquark condensates. While at zero chemical potential 
and finite temperature, magnetic field enhances the condensates, at zero temperature, the 
critical chemical potential decreases with increasing magnetic field.
\end{abstract}

%
%\begin{keyword}
%Covariant regularization, \ spontaneous chiral symmetry breaking, \ PNJL model, \ general spin 0 eight-quark interactions,
%\ finite temperature and chemical potential.
%\PACS 12.38.Mh \ 11.30.Qc \ 71.27.+a \ 12.38-t
%\end{keyword}
\end{frontmatter}

% main text
\section{Introduction}
\label{}
The effect of high density and/or high temperature on the QCD vacuum has been a major theoretical 
and experimental challenge in the physics of strong interaction. In addition to the effects of 
high temperature and density, the effect of high magnetic field has also attracted recent 
attention of physicists recently. The motivation behind this study is the possibilty of creating 
ultra strong magnetic fields in non central collisions at RHIC and LHC which are estimated to be 
of hadronic scale \cite{larrywarringa,skokov} of the order of $eB\sim 2 m_\pi^2$ 
($m_\pi^2\simeq 10^{18}$ Gauss) at RHIC, to about $eB\sim 15 m_\pi^2$ at LHC \cite{skokov}.

The studies of the effect of magnetic field on the vacuum structure of QCD has indicated that magnetic 
field acts as a catalyser of chiral symmetry breaking (CSB) \cite{igormag,miranski}. On the 
other hand, it has been argued recently that there is an inverse magnetic catalysis at finite baryon 
chemical potential \cite{schmitt}.

In this work, we investigate the effect of finite density and temperature on the vacuum structure 
in the context of CSB in a magnetic background using a variational method for 3-flavor NJL model.

\section{Model interaction and gap equation}
We shall consider hot and dense quark matter in a constant magnetic field $\vec B$ in the $z-$ direction 
wich can be obtained from a electromagnetic vector potential given by $A_\mu(\vec x)=(0,0,Bx,0)$. 
Solving the Dirac equation we get the energy of the n-th Lanadu level given as $\epsilon_n^i=\sqrt{m_i^2+p_z^2+2n|q_i|B}\equiv\sqrt{m_i^2+|\vec p_i^2|}$. $q_i$ is the electromagnetic 
charge of the quark of the i-th flavor with current quark mass $m_i$.

To study CSB, we consider a trial state with quark-antiquark pairs as
\begin{equation} 
|\Omega\rangle= \exp\left(\sum_{n=0}^{\infty} \int{d\vec p_{y,z} 
{q_r^i}^\dag(n,\vec p_{y,z})a_{r,s}^i(n,p_z)f^i
(n,\vec p_{y,z})\tilde q_s^i(n,-\vec p_{y,z})} -h.c.\right)|0\rangle.
\label{u0}
\end{equation} 
In the above ansatz for the ground state, $f^i(n,p_z)$ is a real function describing the quark 
antiquark condensates related to the vacuum realignment for chiral symmetry breaking. This will be 
determined from the extremization of the thermodynamic potential. The spin dependent structure 
$a_{r,s}^i$ is given by 
\begin{equation}
a_{r,s}^i=\frac{1}{|\vec p_i|}\left[-\sqrt{2n|q_i|B}\delta_{r,s}-ip_z\delta_{r,-s}\right]
\end{equation}
The ansatz of Eq.(\ref{u0}) is a Bogoliubov transformation of the vacuum state $|0\rangle$. Further 
the effect of temperature and density can be included through a thermal Bogoliubov transformation 
using thermofield dynamics. All these formulations are discussed in detail in reference \cite{bhas}.

The expectation value of the chiral order parameter for the $i$-th flavor for this ground state 
is given by
\begin{equation}
\langle\bar\psi_i\psi_i\rangle_{\beta,\mu} =
-\sum_{n=0}^\infty\frac{N_c|q_i|B\alpha_n}{(2\pi)^2}
\int{dp_z}\cos\phi^i \left(1-n_-^i(k,\beta)-n_+^i(k,\beta)\right)\equiv-I_i,
\label{Ii}
\end{equation}

where, $\alpha_n=(2-\delta_{n,0})$ is the degeneracy factor of the $n$-th Landau level and $n_-$ and 
$n_+$ are distribution functions of quarks and anti quarks respectively. We have defined 
$\phi^i=\phi_0^i-2f_i$ with $\cos\phi_0^i=m_i/\epsilon_{ni}$ and 
$\sin\phi_0^i=|\vec p_i|/\epsilon_{ni}$ \cite{bhas}. 

Now for calculating the thermodynamic potential, we consider the 3-flavor Nambu Jona Lasinio model 
including the Kobayashi-Maskawa-t-Hooft (KMT) determinant interaction term. The corresponding Hamiltonian 
density is given as
\begin{eqnarray}
{\cal H} & = &\psi^ \dagger(-i\vec \alpha \cdot \vec \nabla-qBx\alpha_2
+\gamma^0 \hat m )\psi
-G_s\sum_{A=0}^8\left[(\bar\psi\lambda^A\psi)^2-
(\bar\psi\gamma^5\lambda^A\psi)^2\right]\nonumber\\
&+&K\left[{ det_f[\bar\psi(1+\gamma_5)\psi]
+det_f[\bar\psi(1-\gamma_5)\psi]}\right]
\label{ham}
\end{eqnarray} 
Where $\hat m$=diag$_f(m_u,m_d,m_s)$ is the current quark mass matrix in the flavor space. We 
assume isospin symmetry with $m_u$=$m_d$. $\lambda^A$, $A=1,\cdots 8$ denote the Gellman matrices 
acting in the flavor space and $\lambda^0 = \sqrt{\frac{2}{3}}\,1\hspace{-1.5mm}1_f$, 
$1\hspace{-1.5mm}1_f$ as the unit matrix in the flavor space. The determinant term 
$\sim K$ breaks $U(1)_A$ symmetry. If the mass term is neglected, the overall symmetry is 
$SU(3)_V\times SU(3)_A \times U(1)_V$. This spontaneously breaks to $SU(3)_V \times U(1)_V$ 
implying the conservation of the baryon number and the flavor number. The current quark mass term 
introduces additional explicit breaking of chiral symmetry leading to partial conservation of the 
axial current.

Now minimization of the thermodynamic potential for our model w.r.t the chiral condensate function 
$f _i (p_z)$ leads to the mass gap equation.
\begin{equation}
M_i = m_i + 4GI_i + 2K|\epsilon_{ijk}|I_j I_k
\label{gapeq}
\end{equation}
and the thermodynamic potential is given as
\begin{eqnarray}
\Omega = &-&\sum_{n,i}\frac{N_c\alpha_n|q_iB|}{(2\pi)^2}\int{dp_z\omega_{i}} + 
2G\sum_i I_i^2 + 4KI_1 I_2 I_3 \nonumber\\
 &-& 
\sum_{n,i}\frac{N_c\alpha_n|q_iB|}{(2\pi)^2\beta}\int{dp_z[\ln{\lbrace 1+e^{-\beta(\omega_i-\mu_i)}
\rbrace} + \ln{\lbrace 1+e^{-\beta(\omega_i+\mu_i)}\rbrace}]}
\label{thermtmuB}
\end{eqnarray}
where, $\omega_{i,n}=\sqrt{M_i^2+p_z^2+2n|q_i|B)}$ is the excitation energy with the ``constituent 
quark mass'' $M_i$. The zero temperature and the zero density contribution  of the thermodynamic 
potential ($\Omega(T=0,\mu=0)$) in the above is ultraviolet divergent, which  is also transmitted 
to the gap equation Eq.(\ref{gapeq}). To remove this divergence, we use a regularization procedure 
involving addition and subtraction of a divergent term which is discussed in detail in ref.\cite{bhas}.

\begin{figure}[b]
\centering
\begin{minipage}{\columnwidth}
$\begin{array}{cc}
\includegraphics[width= 0.4 \columnwidth]{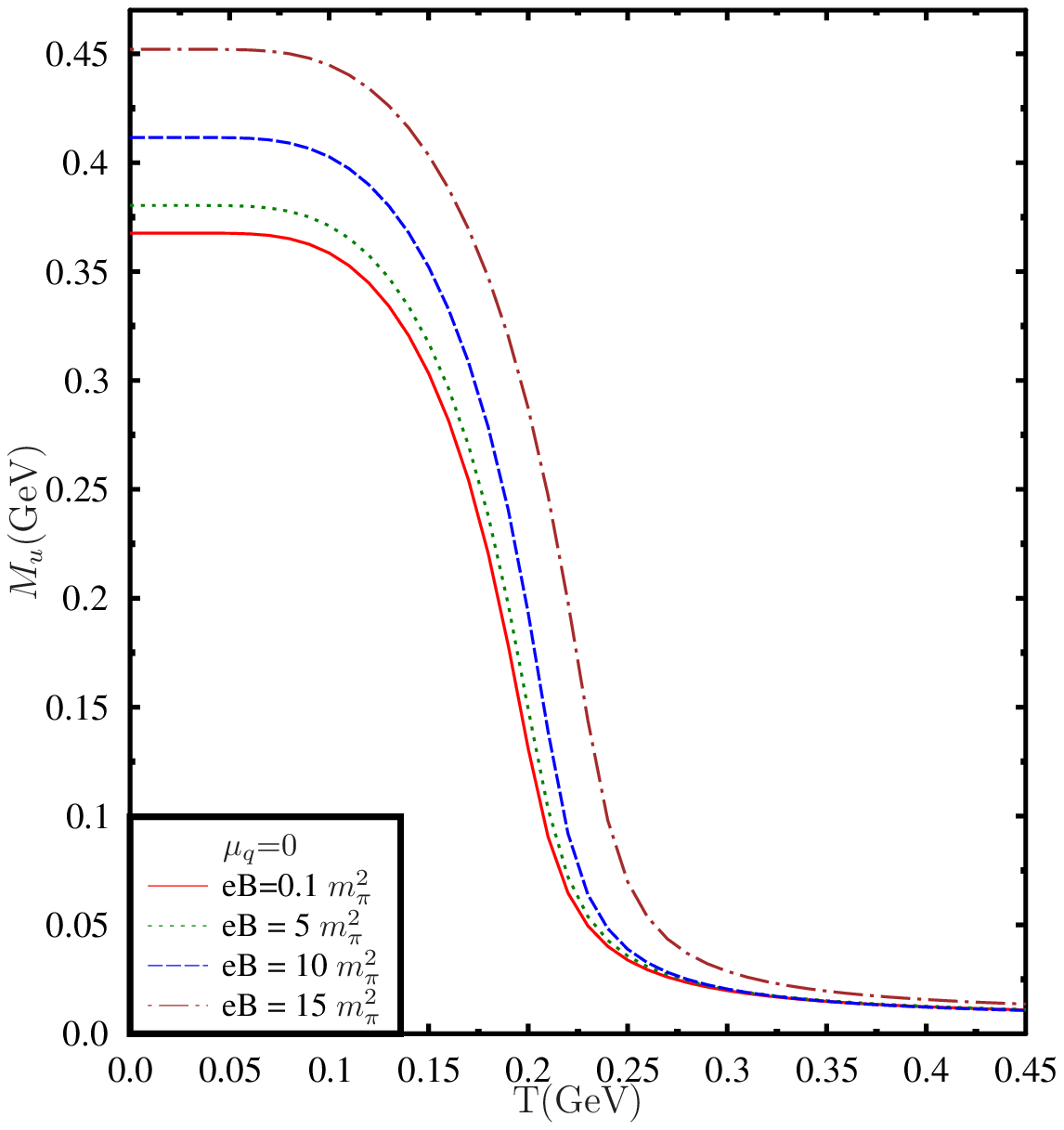}&
\includegraphics[width= 0.4 \columnwidth]{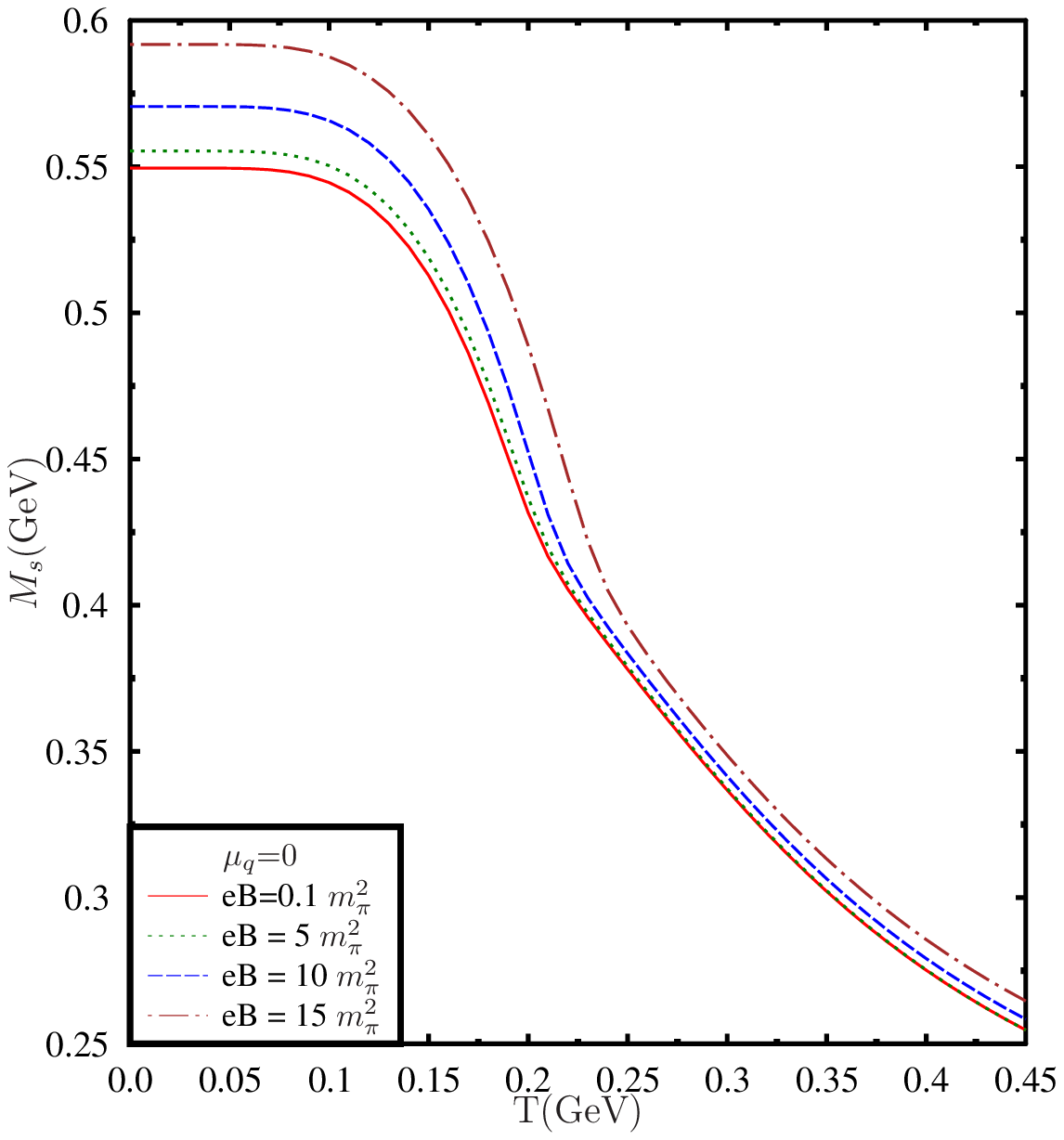}\\
\end{array}$
\end{minipage}
\caption{Constituent quark masses with temperature for different magnetic fields. Figure in the 
left panel shows the mass of up quark $M_u$ at zero baryon chemical potential as a function of 
temperature for different values of the magnetic field. Figure in the right panel shows the same 
for the strange quark mass $M_s$. Both the subplots correspond to nonzero values for the current 
quark masses given as $m_u$=5.5 MeV and $m_s$=140.7 MeV.}
\end{figure}

\section{Results and discussions}
For numerical calculation, we choose the following parametrization of the hamiltonian given in 
Eq.(\ref{ham}). We take $m_u=m_d=5.5$ MeV and $m_s=0.1407$ GeV. The three momentum cut off for 
NJL model is $\Lambda=0.6023$ GeV. The dimensionless couplings are $G_s\Lambda^2=1.835$ and 
$K\Lambda^5=12.36$. With these parameters, the vacuum masses of up and down quarks turn out 
to be 368 MeV and mass of the strange quark is 549 MeV.
In Fig. 1, we have shown the effect of temperature on the constituent quark mass for different 
magnetic field strength at zero baryon chemical potential. The constituent quark mass smoothly 
approaches the current quark mass as temperature is increased. The magnetic field clearly enhances 
the chiral condensate.
In Fig. 2, we  show the effect of magnetic field and chemical potential on chiral symmetry 
breaking. As chemical potential is increased there is a first order transition with the order parameter
changing discontinuously at the critical chemical potential $\mu_c$. Although the
condensate value increase with the magnetic field before the transition, the value of $\mu_c$ {\em decreases} with magnetic field.
This phenomena is termed as inverse magnetic catalysis in Ref.\cite{schmitt} where chiral symmetry breaking 
has been considered
in a holographic model.
A more detailed discussion on the effect of magnetic field, temperature and density on chiral 
symmetry breaking, equation of state as well as charge netral matter has been given in Ref.\cite{bhas}.

\begin{figure}[htp]
\centering
\begin{minipage}{\columnwidth}
$\begin{array}{cc}
\includegraphics[width= 0.38 \columnwidth]{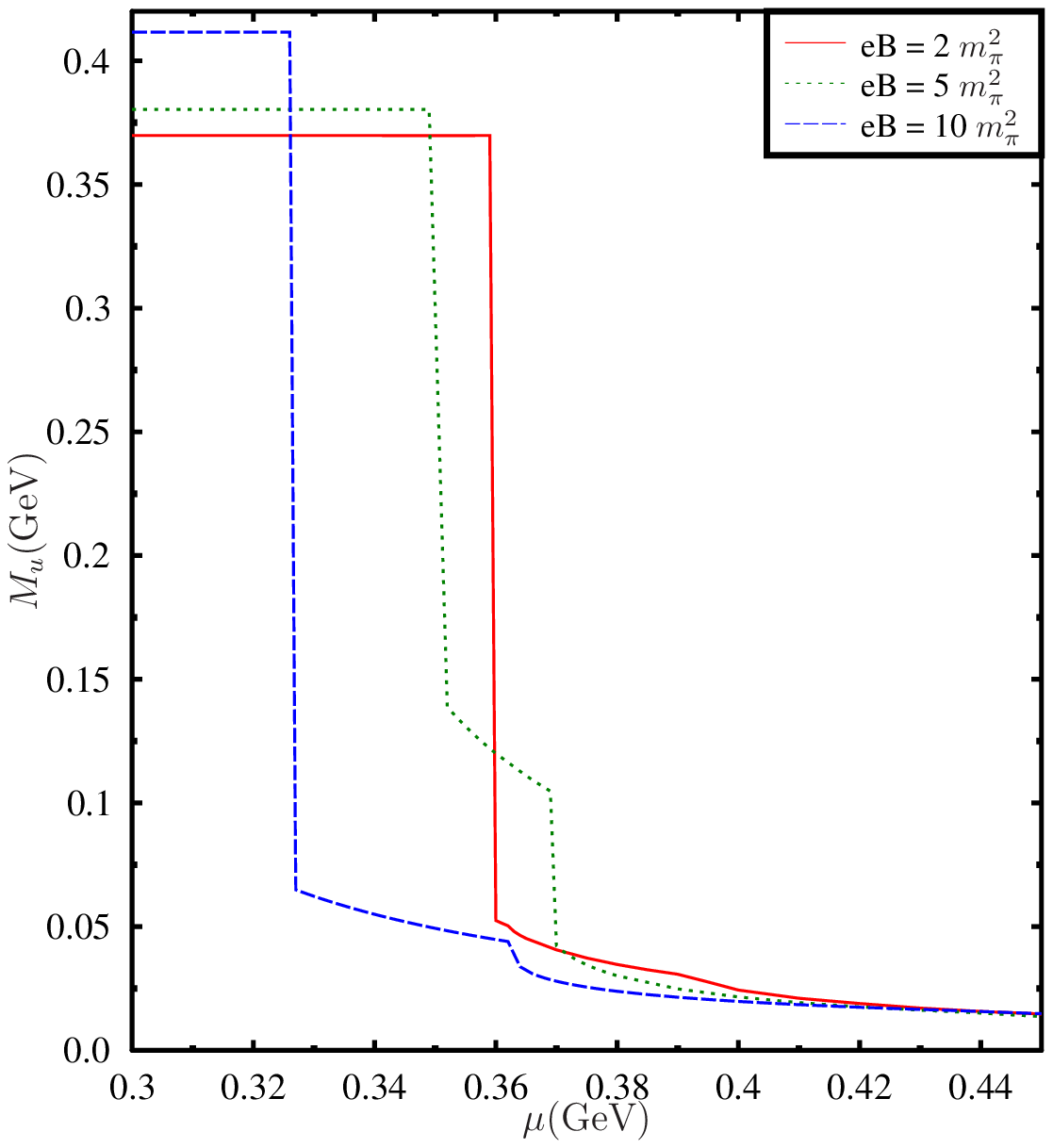}&
\includegraphics[width= 0.38 \columnwidth]{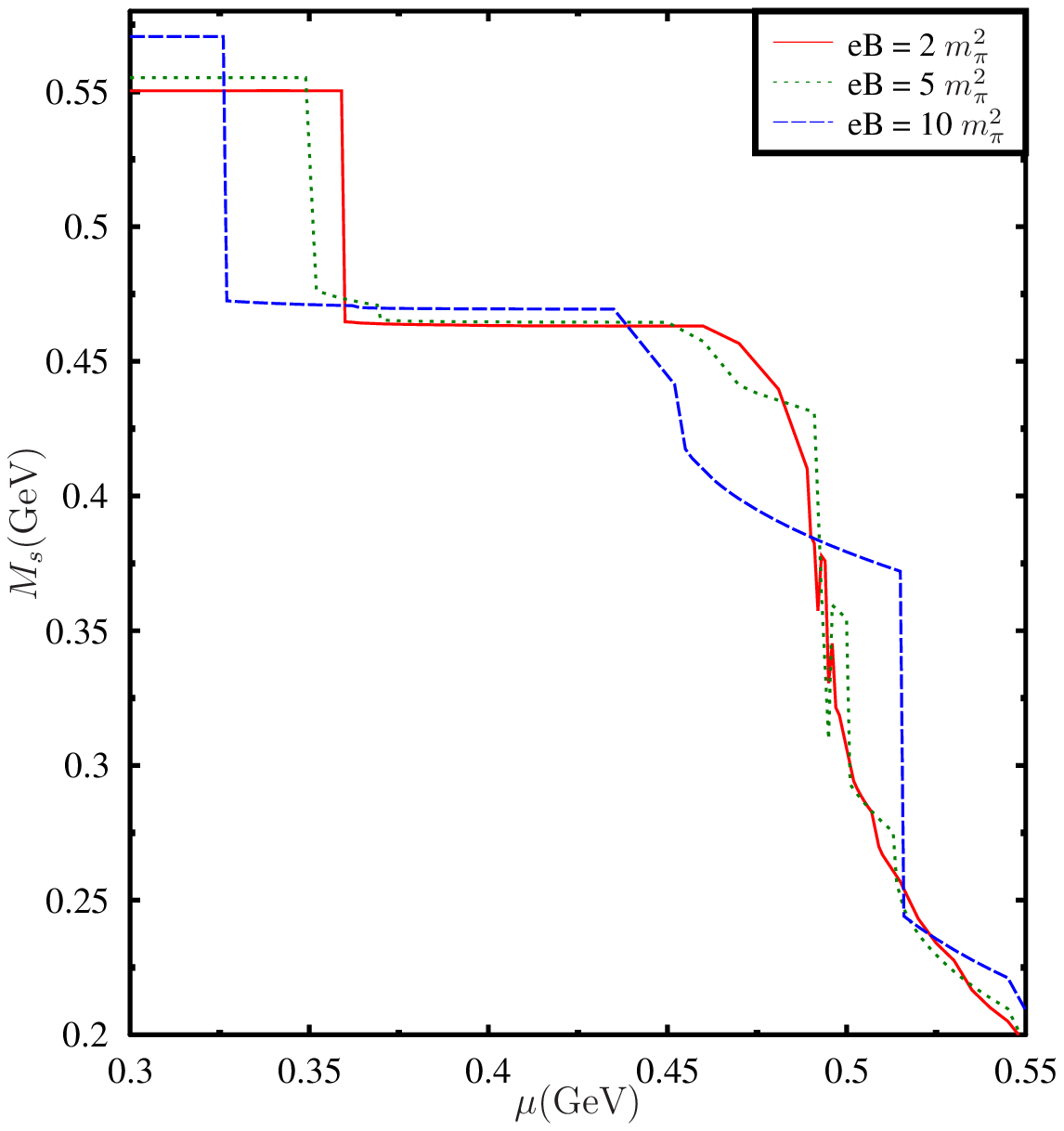}\\
\end{array}$
\end{minipage}
\caption{Constituent quark masses as functions of $\mu_q$ for $T=0$. The up quark masses are 
shown in left panel and figure in the right panel shows strange quark masses for different 
strength of magnetic fields.}
\end{figure}

%%%%%%%%%%%%%%%%%%%%%%%%%%%%%%%%%%%%%%%%%%%%%%%%
%% BACKMATTER
%%%%%%%%%%%%%%%%%%%%%%%%%%%%%%%%%%%%%%%%%%%%%%%%

\def\larrywarringa{D.Kharzeev, L. McLerran and H. Warringa, {\NPA{803}{227}{2008}};
K.Fukushima, D. Kharzeev and H. Warringa,{\PRD{78}{074033}{2008}}.}
\def\skokov{V. Skokov, A. Illarionov and V. Toneev, Int. j. Mod. Phys. A {\bf 24}, 5925,
(2009).}
\def\igormag{E.V. Gorbar, V.A. Miransky and I. Shovkovy,{\PRC{80}{032801(R)}{2009}};
ibid, arXiv:1009.1656[hep-ph].}
\def\miranski{V.P. Gusynin, V. Miranski and I. Shovkovy,{\PRL{73}{3499}{1994}};
{\PLB{349}{477}{1995}}; {\NPB{462}{249}{1996}.}}
\def\schmitt{F. Preis, A. Rebhan, A Schmitt; arXiv:hep-th/1012.4785}
\def\bhas{B. Chatterjee, H. Mishra and A. Mishra; arXiv:hep-ph/1101.0498}

\end{document}